\documentclass[final,5p,times,twocolumn]{elsarticle}

\usepackage{amssymb}
\usepackage{amsthm}

\usepackage{amsmath,amsfonts}
\usepackage{siunitx}
\usepackage{caption}
\usepackage{subcaption}
\usepackage{mathrsfs}
\usepackage{color}
\graphicspath{{./figures/}}
\newcommand{\filt}[1]{\widetilde{#1}}
\newcommand{\rfilt}[1]{\overline{#1}}
\newcommand{\rfluc}[1]{{#1}'}
\newcommand{\fluc}[1]{{#1}''}

\newcommand{\lp}{\left(}
\newcommand{\rp}{\right)}

\newcommand{\pdt}[2][]{\partial_{#1}\lp{#2}\rp}

\newcommand{\pdn}[1]{\nabla\cdot\lp{#1}\rp}
\newcommand{\pdns}[1]{\nabla\lp{#1}\rp}
\newcommand{\pdnl}[1]{\nabla^2\lp{#1}\rp}

\newcommand{\vect}[1]{\mathbf{#1}}

\makeatletter
\DeclareFontEncoding{LS2}{}{\@noaccents}
\makeatother
\DeclareFontSubstitution{LS2}{stix}{m}{n}
\DeclareSymbolFont{largesymbolsstix}{LS2}{stixex}{m}{n}
\DeclareMathDelimiter{\lbrbrak}{\mathopen}{largesymbolsstix}{"EE}{largesymbolsstix}{"14}
\DeclareMathDelimiter{\rbrbrak}{\mathclose}{largesymbolsstix}{"EF}{largesymbolsstix}{"15}

\newcommand{\avgst}[1]{\lbrbrak{#1}\rbrbrak}

\newcommand{\reff}[1]{Fig.~\ref{#1}}

\newcommand{\refe}[1]{Eq.~\eqref{#1}}

\newcommand{\refs}[1]{Sec.~\ref{#1}}

\newcommand{\reft}[1]{Table~\ref{#1}}

\newcommand{\press}[1]{{#1}}
\newcommand{\picsize}{\small}
\newcommand{\picbox}[1]{{#1}}

\journal{ }

\begin{document}

\begin{frontmatter}



\title{Applying Physics-Informed Enhanced Super-Resolution Generative Adversarial Networks to Finite-Rate-Chemistry Flows and Predicting Lean Premixed Gas Turbine Combustors}


\author[jsc,rwth]{Mathis Bode\corref{cor1}}\ead{m.bode@itv.rwth-aachen.de}
\address[jsc]{{J\"ulich Supercomputing Centre, Forschungszentrum J\"ulich GmbH},{52425 J\"ulich},{Germany}}
\address[rwth]{{Fakult\"at f\"ur Maschinenwesen, RWTH Aachen University},{Templergraben 64},{52056 Aachen},{Germany}}

%

\cortext[cor1]{Corresponding author}

\begin{abstract}
The accurate prediction of small scales in underresolved flows is still one of the main challenges in predictive simulations of complex configurations.  Over the last few years, data-driven modeling has become popular in many fields as large, often extensively labeled datasets are now available and training of large neural networks has become possible on graphics processing units (GPUs) that speed up the learning process tremendously. In fact, the successful application of deep neural networks in fluid dynamics, such as for underresolved reactive flows, is still challenging. This work advances the recently introduced PIESRGAN to reactive finite-rate-chemistry flows. However, since combustion chemistry typically acts on the smallest scales, the original approach needs to be extended. Therefore, the modeling approach of PIESRGAN is modified to accurately account for the challenges in the context of laminar finite-rate-chemistry flows. The modified PIESRGAN-based model gives good agreement in a priori and a posteriori tests in a laminar lean premixed combustion setup. Furthermore, a reduced PIESRGAN-based model is presented that solves only the major species on a reconstructed field and employs PIERSGAN lookup for the remaining species, utilizing staggering in time. The advantages of the discriminator-supported training are shown, and the usability of the new model demonstrated in the context of a model gas turbine combustor.
\end{abstract}




\end{frontmatter}


\section{Introduction}
Data-driven methods have evolved as important tools in many applications. With the growth in availability and size of often extensively labeled datasets, data-driven models have potential to become more and more accurate. Recently, significant advances in terms of functionality and processing with optimized graphics processing units (GPUs), such continuously updating network weights in a data-fed training process to minimize loss functions by machine learning (ML) and deep learning (DL), have been made. Examples for successful applications cover a broad range including image processing~\cite{wang2019,greenspan2016guest}, speech recognition~\cite{hinton2012deep}, acceleration of drug developments~\cite{bhati2021}, and learning of optimal complex control~\cite{Vinyals2019}. Some of these examples employ generative adversarial networks (GANs)~\cite{goodfellow2014generative} and demonstrate that GANs are one powerful way to use DL. A GAN consists of two DL networks: a generator network, which is finally used for data generation, and a discriminator network supporting the training process by assessing if data is real or generated by the generator. As the generator attempts to "fool" the discriminator and the discriminator becomes better and better in discrimination of real and generated data with increasing training time, the overall quality of the generated data improves. Both parts can be independently updated during the training process and are coupled by the adversarial loss term. By design, GANs learn unsupervised from unlabeled data, which is a significant advantage for many physical applications, for which available training data is often sparse and does not cover the entire target parameter space of interest.

Many recent applications of data-driven models can also be found in the field of fluid dynamics, such as the usage in simulations requiring modeling of the small scales, in particular Reynolds-averaged Navier-Stokes (RANS) simulations and large-eddy simulations (LESs) \cite{ihme2009,maulik2017neural,bode2018,srinivasan2019predictions,lapeyre2019training,fukami2019synthetic,bode2019bspline,dalessio2020}. 
This includes the idea of physics-informed networks~\cite{raissi2019}, which has recently emerged. Physics-informed networks employ architectures or loss functions that are designed to support known properties of the underlying physical problems to regularize the learning procedure. Kutz~\cite{kutz2017} summarized more applications of ML/DL in the field of fluid mechanics.

Bode et al.~\cite{bode2019,bode2021dad,bode2022spray} introduced physics-informed enhanced super-resolution generative adversarial networks (PIESRGANs) as subfilter closure for LES. They advanced ESRGANs (enhanced super-resolution GANs), which were originally developed for two-dimensional (2-D) image super-resolution~\cite{wang2018esrgan}, to handle 3-D physical data and combined it with physical information during the training process. The general training procedure of this network relies on high-fidelity data ("H"), e.\,g., from a fully resolved direct numerical simulation (DNS), which are filtered ("F") as part of training. The combination of fully resolved data $\Phi_\mathrm{H}$ and filtered data $\Phi_\mathrm{F}$ is used to train the full network with the goal to be able to achieve as accurate as possible reconstructed data ("R") $\Phi_\mathrm{R}$ with the filtered data as input, which can be used to close any required subfilter terms. It was shown that PIESRGAN gives very good results for LES of turbulence~\cite{bode2021dad,gauding2021,bode2022pdl} and of spray combustion based on the multiple representative interactive flamelet model (MRIF)~\cite{bode2021dad,bode2022dlc} if it is used for modeling the unclosed velocity and scalar mixing terms in the filtered Navier-Stokes equations. However, PIESRGAN has not been applied to obtain the filtered chemical source terms directly, which is done in this work for the first time on the example of laminar lean premixed combustion. Even though LES is not necessary in this case due to the lack of turbulence, the challenge to accurately predict small scales from filtered fields remains comparable, as the prediction of flame dynamics and pollutant formation requires fine grids even in the case of laminar flows. 

Power plants with gas turbines are an important pillar for combating climate change, as they combine multiple advantages. They have low start-up times and can quickly follow the load demand, balancing the volatility of wind and solar power generation. The investment costs are relatively low, which is important for peaker or backup plants which are only operated for a low number of hours per year. And finally, natural gas is a much cleaner power source than coal, with an easy pathway to incorporate renewable hydrogen or e-methane in the future. One important challenge in the development of advanced highly efficient gas turbine combustors is the minimization of CO emissions under low part load conditions. In particular, interaction between flames and cold secondary air from unfired stages or combustor cooling can lead to perturbations of the oxidation layer in premixed flames resulting in incomplete CO burnout. Modeling of such interaction is often difficult with classical combustion models such as flamelet models, or very expensive when finite rate chemistry is applied. Thus, this physical problem appears to be a good application for advanced, data-driven models such as PIESRGAN. 

In this work, a simplified laminar lean premixed model configuration, representing the underlying physical processes described above, is chosen as a first target case for the development of PIERSGAN for reactive finite-rate-chemistry flows.
As the scope of this work is on modeling finite-rate-chemistry flows with PIESRGAN, the momentum fields are still computed on DNS resolution, and the full integration in PIESRGAN is left for future work.

\section{Case setup of simplified lean premixed combustion}
To investigate combustion in stationary gas turbines, L\"uckerath et al.~\cite{lueckerath2011} developed a test rig with staged combustors. Their configuration features 8 circumferentially arranged burners around a central pilot flame and can be run such that not all burners are loaded. Loaded burners then interact with neighboring air jets with the impact on chemistry, especially CO, depending on the mixing velocity. This setup is mimicked with two different cases in this work: a laminar setup and a turbulent setup. All cases in this work used methane as fuel, and a chemical mechanism derived from GRI3.0~\cite{smith_gri-mech_nodate} by Luca et al.~\cite{luca2017} with 16 species and 72 reversible reactions was employed.

\begin{figure*}[h!]
\picsize
\centering
        \begin{subfigure}[b]{\textwidth}
            \centering
            \picbox{\includegraphics[width=\textwidth]{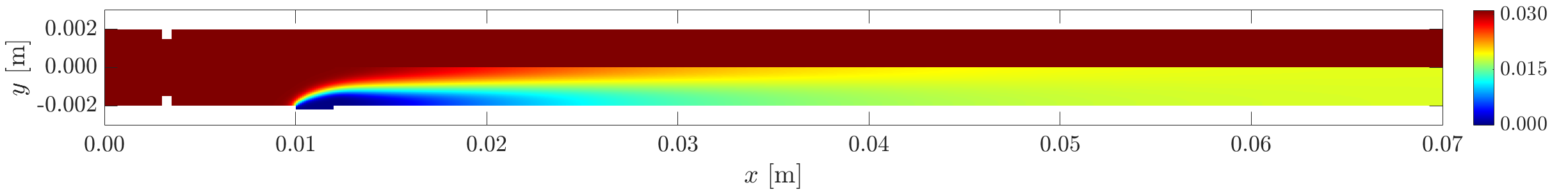}}
            \caption{$Z$ [-]}    
            \label{dpl:sfig:ov:VZ}
        \end{subfigure}
        \vskip 1mm

        \begin{subfigure}[b]{\textwidth}
            \centering
            \picbox{\includegraphics[width=\textwidth]{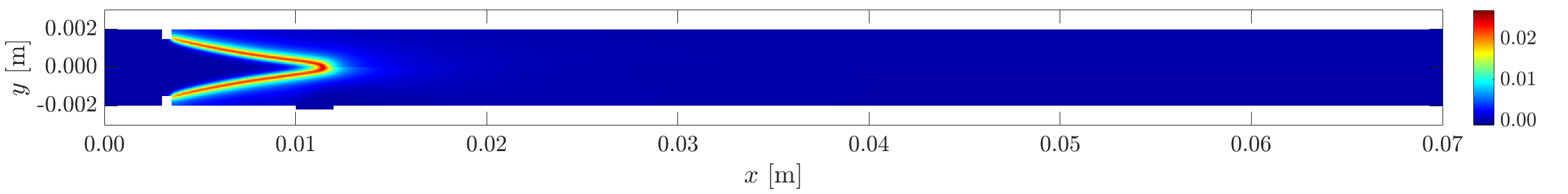}}
            \caption{$Y_\mathrm{CO}$ [-]}    
            \label{dpl:sfig:ov:VY}
        \end{subfigure}
        \vskip 1mm

        \begin{subfigure}[b]{\textwidth}
            \centering
            \picbox{\includegraphics[width=\textwidth]{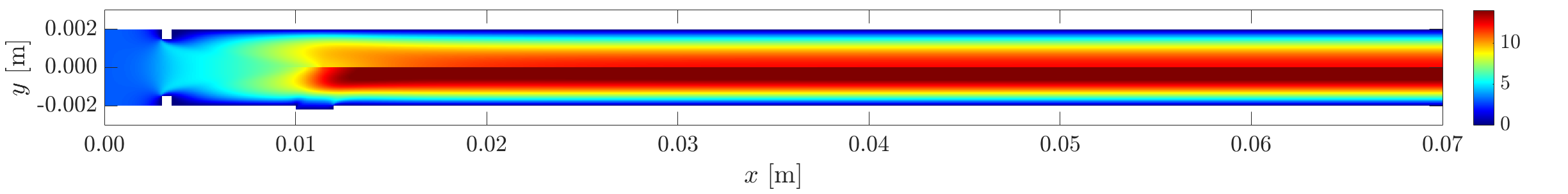}}
            \caption{$U$ $[\mathrm{m}/\mathrm{s}]$}    
            \label{dpl:sfig:ov:VU}
        \end{subfigure}
        \vskip 1mm

          \begin{subfigure}[b]{\textwidth}
            \centering
            \picbox{\includegraphics[width=\textwidth]{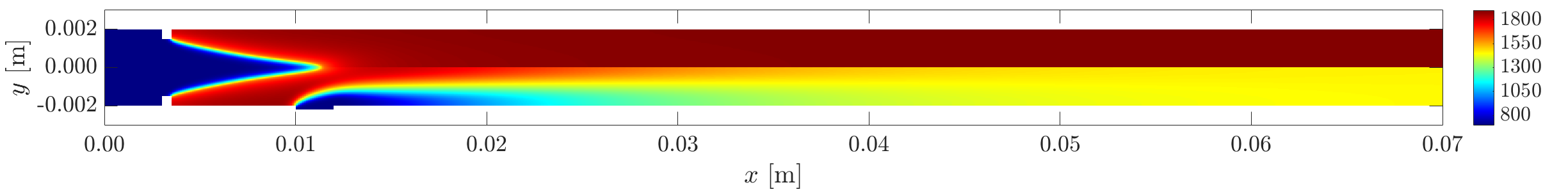}}
            \caption{$T$ $[\mathrm{K}]$}    
            \label{dpl:sfig:ov:VT}
        \end{subfigure}
\caption{Fully resolved DNS results of the laminar model setup for the mixture fraction, CO mass fraction, axial velocity, temperature for the case with wall, i.\,e., $V_\mathrm{inlet}=\SI{0.0}{\meter\per\second}$, (top) and $V_\mathrm{inlet}=\SI{4.0}{\meter\per\second}$ (bottom). Note that not the full length in $x$-direction is shown.}
\label{dpl:fig:ov}
\end{figure*}

%
A simplified configuration to reproduce the main characteristics of a laminar flame/air jet interaction and the resulting increase of CO was employed in this work.  The focus is on one premixed fuel inlet, and the flame is diluted with air by two symmetric side inlets downstream. To keep simulation cost low, this is modeled as a 2-D model domain discretized by a uniform mesh and periodic boundary conditions in the third dimension, and the flame was resolved by 18 grid points per flame thickness, resulting in a $[4096\times 192]$ mesh. Even though the setup considered here is symmetric, the axis symmetry was not used in the simulations and a domain with two separate inlets was considered to enable the possibility to study unsymmetric configurations with more complex dilution patterns in the future with the same setup. The inlet velocity $U_\mathrm{inlet}$ was chosen as \SI{3.0}{\meter\per\second}, the air-fuel equivalence ratio $\phi$ was set to \num{0.55}, and the inlet velocities of the side inlets $V_\mathrm{inlet}$ were varied from \SI{0.0}{\meter\per\second} (reference case without dilution) to \SI{8.0}{\meter\per\second}. For training the network, the cases with $V_\mathrm{inlet}=\SI{0.0}{\meter\per\second}$ and $V_\mathrm{inlet}=\SI{4.0}{\meter\per\second}$ were used, which are also illustrated in \reff{dpl:fig:ov} with their respective mixture fraction fields $Z$, CO mass fraction fields $Y_\mathrm{CO}$, axial velocity fields $U$, and temperature fields $T$.

This laminar lean premixed combustion application is used as target case for the development of PIESRGAN for reactive finite rate chemistry flows for three reasons. First, the laminar case allows to fully focus on the proper reconstruction of the finite rate chemistry scalar fields without dealing with the issue of turbulent fluctuations. Thus, the velocities were not used for training in this work; however, note that this does not mean that the velocities are not considered by the network. As the chemistry is coupled with the velocities, the effect of different flow velocities and potentially related different strain rates is implicitly recognized by the network by recognizing different scalar field configurations. Second, models for premixed combustion with a focus on emissions at varying mixtures often lack accuracy if the species concentration exceeds its common range at the local mixture. Third, the resulting chemistry of lean premixed combustion cases can be highly non-linear even for relatively simple configurations, making it a very appealing target case for the developed model. 

The goal of such a simulation could be to accurately determine CO emissions at different levels of interaction between the flame and the lateral inflow of air. Furthermore, the underlying physical processes leading to different CO levels at the exhaust at varying levels of interaction are interesting. The leading quantity to compare cases has been the average mass flux of CO in the domain. It will be found as part of the variations in the application chapter that CO emissions in the lean yet unperturbed configuration converged quickly to the equilibrium concentration. Mild dilution from the lateral inlet reduces the CO emissions below the unperturbed case, without the configuration reaching its equilibrium concentration. Even stronger dilution with air further reduces the equilibrium concentration, but CO consumption turns out to be very slow, with a virtual freeze of the CO concentration at one point resulting in an overall increase of CO emissions at the outlet. Above-mentioned average mass of CO in the domain will be used as key indicator for the accuracy of the modeling approach introduced here. Furthermore, the full transition phase from the initialization to steady-state is used for training and validation, although the final result is steady-state. This increases the number of learning samples and is also more challenging for the model, as multiple ”operating points” are passed during initialization.


\section{PIESRGAN}
This section describes the PIESRGAN model for finite-rate-chemistry flows. It starts with a summary of PIESRGAN and develops the extensions for finite-rate chemistry afterward step-by-step.

\subsection{Base PIESRGAN model}
\begin{figure*}[htbp]
\centerline{\includegraphics[width=\textwidth]{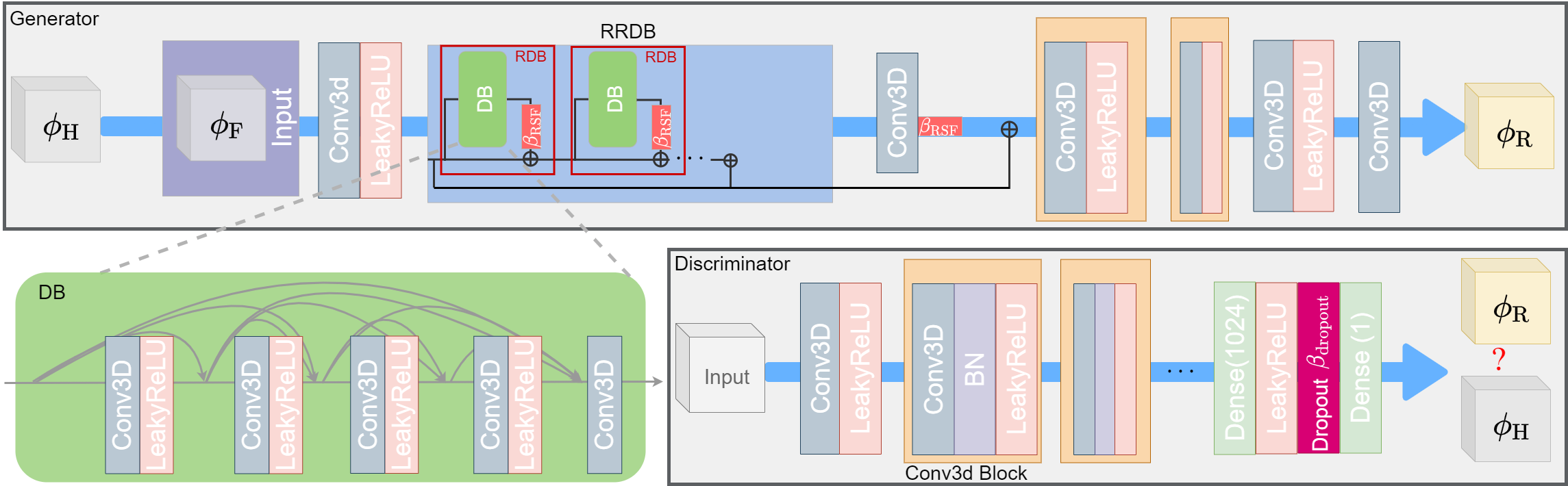}}
\caption{Sketch of PIESRGAN. "H" denotes high-fidelity data, "F" are corresponding filtered data, and "R" are the reconstructed data. The components are: Conv3D - 3-D Convolutional Layer, LeakyReLU - Activation Function, DB - Dense Block, RDB - Residual Dense Block, RRDB - Residual in Residual Dense Block, $\beta_\mathrm{RSF}$ - Residual Scaling Factor, BN - Batch Normalization, Dense - Fully Connected Layer, Dropout - Regularization Component, and $\beta_\mathrm{dropout}$ - Dropout Factor. Image from \cite{bode2021dad}.}
\label{dpl:fig:piesrgan}
\end{figure*}

PIESRGAN for turbulence and passive scalar transport was introduced by Bode et al.~\cite{bode2021dad}. The method relies on a GAN, which is trained with "H"/"F" data pairs, and a closure algorithm. The suggested network by Bode et al.~\cite{bode2021dad} is sketched in \reff{dpl:fig:piesrgan}. The generator heavily uses combinations of leaky rectified linear unit (LeakyReLU) layers for activation \cite{maas2013} and 3-D CNN layers (Conv3D) \cite{krizhevsky2012imagenet}. Previous architectures relied on the residual block (RB), which contains fundamental architectural elements such as residual dense blocks (RDBs) with skip-connections. This was extended for ESRGAN to the residual in residual dense block (RRDB) with residual scaling factor $\beta_{\mathrm{RSF}}$, helping to avoid instabilities in the forward and backward propagation. The RRDB is essential for the performance of state-of-the-art super-resolution.  The RDBs use dense connections inside, i.\,e., the output from each layer within the dense block (DB) is sent to all the following layers. The discriminator mainly employs basic CNN layers (Conv3D) with LeakyReLU layers for activation with and without batch normalization (BN). The final layers combine a dropout with dropout factor $\beta_{\mathrm{dropout}}$ with a fully connected layer and LeakyReLU activation.

As example to summarize the closure algorithm, the transport of a passive specific scalar $\phi$ over time $t$ in a non-reactive, incompressible flow with constant molecular diffusivity $D$ and without body forces can be described by
\begin{equation}
	\pdt[t]{\phi} + \vect{u}\cdot\pdns{\phi} = D\pdnl{\phi}
	\label{dpl:eq:sca}
\end{equation}
with bold notation for vectors, $\nabla$ as del operator, corresponding to spatial coordinates, such as $\vect{x}=(x_1,x_2,x_3)^\intercal$ for a Cartesian coordinate system or $\vect{x}=(r,\theta,\phi)^\intercal$ for a spherical coordinate system, and $\partial_t$ as time derivative. 

For LES, an homogeneous filter operation, which moreover is assumed to be a Reynolds operator and denoted with an overbar, and Reynolds splitting as
\begin{equation}
  \label{dtr:eq:split}
  \{\cdot\} = \rfilt{\{\cdot\}} + \rfluc{\{\cdot\}}
\end{equation}
with the overbar denoting the filtered part and the prime the subfilter part of a quantity with 
\begin{equation}
  \label{dtr:eq:id1}
  \rfilt{\rfilt{\{\cdot\}}} = \rfilt{\{\cdot\}}\rfilt{1} = \rfilt{\{\cdot\}}
\end{equation}
and
\begin{equation}
  \label{dtr:eq:id2}
  \rfilt{\rfluc{\{\cdot\}}} = 0
\end{equation}
are defined. Introducing Reynolds splitting for $\vect{u}$ and $\phi$ in \refe{dpl:eq:sca} and applying the filter operation to all terms, results in the filtered equation solved in LES. This transport equation reads
\begin{equation}
	\pdt[t]{\rfilt{\phi}} + \rfilt{\vect{u}}\cdot\pdns{\rfilt{\phi}} = D\pdnl{\rfilt{\phi}} - \rfilt{\rfluc{\vect{u}}\cdot\pdns{\rfluc{\phi}}}.
	\label{dpl:eq:sfilt}
\end{equation}
The last term on the right side is unclosed as it requires subfilter information, which are not solved for in LES. This closure is done with PIESRGAN. If $\Phi_\mathrm{F}^n$ denotes a discretized filtered solution at time step $n$, including velocity and all scalar fields, the resulting simulation workflow for closing unclosed terms $\Psi_\mathrm{F}^n$ is as follows:
\begin{enumerate}
\item Use the PIESRGAN to reconstruct $\Phi_\mathrm{R}^n$ from $\Phi_\mathrm{F}^n$.
\item Use $\Phi_\mathrm{R}^n$ to estimate the unclosed terms $\Psi_\mathrm{F}^n$ by evaluating the local terms with $\Phi_\mathrm{R}^n$ and applying a filter operator.
\item Use $\Psi_\mathrm{F}^n$ and $\Phi_\mathrm{F}^n$ to advance the filtered equations to $\Phi_\mathrm{F}^{n+1}$.
\end{enumerate}

\subsection{PIESRGAN for lean premixed combustion}
The application of finite rate chemistry usually comes along with the solution of coupled transport equations for the mass fractions $Y$ of every species $alpha$. This set of equations reads
\begin{equation}
	\pdt[t]{\rho Y_\alpha} + \pdn{\rho \vect{u} Y_\alpha} = \pdn{\rho D_\alpha \pdns{Y_\alpha}} + \dot{\omega}_\alpha
	\label{dpl:eq:rsca}
\end{equation}
with $\dot{\omega}_\alpha$ as chemical source term of species $\alpha$, which can be also written as $\dot{\omega}_\alpha=\rho S_\alpha$. All equations are coupled by the chemical mechanism, which provides $D$ and $\dot{\omega}_\alpha$ depending on the local composition of all species as well as temperature and pressure. 

In contrast to \refe{dpl:eq:sca}, $\rho$ and $D$ are not constant for reactive flows. Due to the variable density, Favre-filtering is therefore more convenient in these cases compared to the earlier introduced Reynolds-filtering. Favre-filtering is denoted by a tilde and defined as
\begin{equation}
 \filt{\{\cdot\}} = \frac{\rfilt{\{\rho \cdot\}}}{\rfilt{\rho}}.
\end{equation}
The corresponding Favre-decomposition reads
\begin{equation}
\{\cdot\} = \filt{\{\cdot\}} + \fluc{\{\cdot\}}.
\end{equation}
Note that $\rfilt{\fluc{\{\cdot\}}} \neq 0$ but $\rfilt{\fluc{\rho\{\cdot\}}} = 0$.

Introducing Favre-decomposition in \refe{dpl:eq:rsca}, using the continuity equation, and Reynolds-filtering all terms, leads to the filtered set of equations as
\begin{equation}
	\rfilt{\rho} \pdt[t]{\filt{Y}_\alpha} + \rfilt{\rho} \filt{\vect{u}} \cdot \pdns{ \filt{Y}_\alpha} = \rfilt{\pdn{\rho D_\alpha \pdns{Y_\alpha}}} -\pdn{\rfilt{\rho}\filt{\vect{u}}\fluc{Y}_\alpha} + \rfilt{\rho}\filt{S}_\alpha.
	\label{dpl:eq:frsca}
\end{equation}
Note that all terms on the right side are unclosed.  While the second term on the right side, the turbulent transport, is similar to the unclosed term in \refe{dpl:eq:sfilt}, the last term, the averaged source term, and the first term, the molecular transport, which is unclosed due to the non-constant diffusion coefficients and the variable density, are new. However, also the new unclosed terms can be closed on the reconstructed fields, however, not as simple as the turbulence term as outlined in this work.

Two important challenges in this context should be highlighted. First, the chemical source terms are often only active in a very fine zone, acting on the smallest scales. This makes it very challenging to accurately model their impact in filtered equations. Second, diffusion models, such as the Curtiss-Hirschfelder approximation \cite{hirschfelder1964}, are often very complex and further effects, such as preferential diffusion, exacerbate the situation even further.

\subsection{Full PIESRGAN model}
\label{dpt:ssec:full}
The target case in this study is laminar, and thus an LES is not necessary for predicting the velocity field. However, filtered equations are still used to model the effect of an underresolved flow field on chemistry.  The effect of such underresolved species fields, as would be commonly encountered in LES or RANS simulation of more complex setups, can be clearly seen in \reff{dpl:fig:conservation}, where the domain-averaged mass of CO deviates by \SI{19}{\%} from the fine DNS grid to the coarser underresolved grid. To keep things simple, a box filter is employed for all fields but the model closure is only used for the scalar fields. Due to the complexity on the finest scales, which need to be reconstructed and the complex physical processes there, a straight-forward usage of the approach by Bode et al.~\cite{bode2021dad} was not successful, resulting in significant deviations in a priori tests and divergence in a posteriori scenarios. As a result, a modified way, in which the full transport equations are solved on the reconstructed grid, is pursued here, retaining the complex direct interaction of transport and chemical source terms. The resulting model can be seen as solving the governing equations on multiple grids, utilizing the super-resolution capabilities of PIERSGAN to achieve the necessary resolution for accurate chemistry computations while solving the overall flow field on a much coarser mesh after mapping the solution back.  The algorithm starts with the filtered solution $\Phi_\mathrm{F}^{n}$ at time step $n$, which includes the entirety of all relevant fields in the simulation, and consists of repeating the following steps:
\begin{enumerate}
\item Use the PIESRGAN to reconstruct $\Phi_\mathrm{R}^n$ from $\Phi_\mathrm{F}^n$.
\item Use $\Phi_\mathrm{R}^n$ to update the chemistry to $\Phi_\mathrm{R}^{n;\mathrm{update}}$ by solving the unfiltered transport equations~(\refe{dpl:eq:sca}) on the mesh of $\Phi_\mathrm{R}^n$.
\item Use $\Phi_\mathrm{R}^{n;\mathrm{update}}$ to estimate the unclosed terms $\Psi_\mathrm{F}^n$ in the filtered transport equations of $\Phi$ by evaluating the local terms with $\Phi_\mathrm{R}^{n;\mathrm{update}}$ and applying a filter operator.
\item Use $\Psi_\mathrm{F}^n$ and $\Phi_\mathrm{F}^n$ to advance the filtered transport equations of $\Phi$ to $\Phi_\mathrm{F}^{n+1}$.
\end{enumerate}
Within this algorithm, the chemistry source terms on the resolved field $\Phi_\mathrm{R}^{n;\mathrm{update}}$ in step~2 can be evaluated by multiple methods, such as tabulated flamelet and finite-rate-chemistry approaches. All approaches benefit from the locally high resolution on the reconstructed mesh. As in other simulations, flamelet approaches might lack accuracy due to their low-dimensional mapping of the thermochemical parameter space, and finite-rate-chemistry techniques can be expected to give more reliable results. Therefore, this work focuses on finite-rate chemistry in the following. However, using tabulated flamelet models might enable even further computing cost reductions for more complex cases. 

\subsection{Implementation details}
As the PIERSGAN is applied to the species fields instead of the velocity field as in \cite{bode2021dad}, the continuity loss term, which was used previously to enforce the continuity equation, is replaced by a physically motivated term enforcing all species to sum up to one, i.\,e., $\sum_i Y_i = 1$.  This condition can also be enforced by computing one mass fraction as one minus the sum of all others, as it is often done in simulations, however, this approach gave worse results in tests in the context of PIESRGAN and was therefore dropped.

With the new term for species conservation, $\beta_4 L_\mathrm{species}$, the used loss function reads
\begin{equation}
\mathcal{L} = \beta_1 \press{L_\mathrm{adv.}} + \beta_2 L_\mathrm{pixel} + \beta_3 L_\mathrm{gradient} + \beta_4 L_\mathrm{species},
\end{equation}
where $\beta_1$, $\beta_2$, $\beta_3$, and $\beta_4$ are coefficients weighting the different loss term contributions; in this work, these coefficients were always equally scaled such that the sum of all non-zero weighting coefficients remained equal to one. Note that all loss terms are non-dimensional, since all operators and input fields used are non-dimensionalized. $L_\mathrm{adversarial}$ is the discriminator/generator relativistic adversarial loss~\cite{wang2019}, which reflects both how well the generator is able to generate high-resolution turbulence samples that look like real, DNS-obtained fully resolved scalar fields, and how well the discriminator is able to distinguish between real and generated data. The pixel loss $L_\mathrm{pixel}$ and the gradient loss $L_\mathrm{gradient}$ are defined as the mean-squared error (MSE) of the quantity itself and of the gradient of the quantity, respectively~\cite{bode2019}. Note that an element conservation term was also elaborated during model development. For the here considered cases, this additional term showed negligible impact on the results and was thus dropped.

Filter type and filter width are also critical implementation details. Bode et al.~\cite{bode2021dad} pointed out that PIESRGAN-LES is very robust with respect to the filter width. Bode~\cite{bode2022dnt} demonstrated how multiple filter widths can be considered during training to achieve flexibility with respect to training mesh size and local filter width. All common filter types can be generally employed with PIESRGAN.

Bode et al.~\cite{bode2021dad} emphasized the importance of proper normalization of input quantities. For the mass fraction fields this was not necessary, as they are non-dimensional by nature. However, other quantities, such as the density, were scaled between zero and one for training and retrieval. Furthermore, it was tried to improve the accuracy by logarithmic scaling of the mass fraction fields to account for their different orders of magnitudes. This did not give a reasonable improvement and therefore was not further pursued. The numerical solver and hyperparameters, such as network parameters learning rates for the generator training and discriminator training, $l_{\mathrm{generator}}$ and $l_{\mathrm{discriminator}}$, were chosen similarly as by Bode et al.~\cite{bode2021dad} and are also summarized in \reft{dpl:tab:hp}.
\begin{table}[!htb]
 \centering
 \caption{Overview of the PIESRGAN hyperparameters. The given ranges represent the sensitivity intervals with acceptable network results. The central values were finally used in this work. Note that the given boundaries are not combined sets of hyperparameters and therefore do not sum up to one.}
\label{dpl:tab:hp}
\begin{tabular}{c c}
\hline
$\beta_1$ & $[\num{0.2e-5},\num{0.6e-4},\num{0.8e-4}]$ \\
$\beta_2$ & $[\num{0.79327},\num{0.86994},\num{0.91812}]$ \\
$\beta_3$ & $[\num{0.04},\num{0.06},\num{0.15}]$ \\
$\beta_4$ & $[\num{0.01},\num{0.07},\num{0.08}]$ \\
$\beta_\mathrm{RSF}$ & $[\num{0.1},\num{0.2},\num{0.3}]$ \\
$\beta_\mathrm{dropout}$ & $[\num{0.2},\num{0.4},\num{0.5}]$ \\
$l_\mathrm{generator}$ & $[\num{1.2e-6},\num{4.5e-6},\num{5.0e-6}]$ \\
$l_\mathrm{discriminator}$ & $[\num{4.4e-6},\num{4.5e-6},\num{8.5e-6}]$ \\
\hline
 \end{tabular}
\end{table}

\subsection{Reduced PIESRGAN model}
%
%
The algorithm introduced in \refs{dpt:ssec:full} raises the question, why the equations are not directly solved on the finer mesh, omitting the costly reconstruction step. In fact, the previously described "full" PIERSGAN model should be seen as idealized proof of concept, establishing that arbitrary super-resolution of reactive flow fields between meshes is possible and yields satisfactory results. To actually achieve a speed-up compared to a conventional, fully resolved simulation, additional modeling steps are necessary. Some possibilities for this are discussed in the following. 

A first trivial optimization is to only reconstruct the fine mesh in cells which are expected to feature high mass fraction gradients, i.\,e., in the vicinity of the flame front. This can be simply tracked by, e.\,g., the progress variable gradient and already reduces the overall cost significantly. The resulting model can then be considered as a form of highly flexible, local and temporal adaptive mesh refinement - which can also be applied on structured grids, where such local refinement otherwise requires complex methods such as overset grids.
An issue that remains is that the computational effort for the reconstruction of species scales quadratically with the number of simultaneously reconstructed quantities. Even if only performed locally on parts of the mesh, this can quickly become technically challenging even on modern GPUs, if large chemical reaction mechanisms are employed. To counteract this effect, a second optimization to significantly reduce the number of reconstructed species has been developed.  For that, the species are split into primary species ("p") and secondary species ("s"). The fine-scale equations are only solved for the primary species while the chemistry evolution of the secondary species is directly predicted by the network based on all other species. This can be seen as a chemtable lookup utilizing the entire chemical space instead of a low-dimensional mapping of it as in traditional chemtable models. It was found that the overall performance of this model and especially the prediction of the evolution of the secondary species can be significantly improved by introducing time staggering for primary and secondary species. Thus, for the presented results with the reduced PIERSGAN model, also denoted $\mathrm{PIESRGAN}_\mathrm{S}$, the velocities were updated in the same time step as the secondary species, while the update of the primary species was 
shifted by half a time step. More precisely, the lookup of the secondary species $Y_{\mathrm{s};i}^{n+1}$ at time $t+1$ by means of PIESRGAN was done with the input quantities $Y_{\mathrm{s};i}^{n}$ and $Y_{\mathrm{p};i}^{n+1/2}$, representing the full thermochemical state space as temperature and density are implicitly included for the case without heat losses considered.

As the full PIESRGAN is a full representation of the reduced version $\mathrm{PIESRGAN}_\mathrm{S}$, it can be used to systematically identify primary and secondary quantities employing AutoML~\cite{hutter2019automated}. Simply put, the AutoML algorithm systematically checks the sensitivity of changes to the primary and secondary species set to find reduced versions with sufficient accuracy. The full PIESRGAN acts as the optimal solution within this optimization and therefore helps with faster convergence. Technically, the optimization of the species set is treated as a hyperparameter optimization, and consequently three new hyperparameters were introduced. First, two new hyperparameters describing the primary and secondary species set, $\vect{\beta}_\mathrm{p}$ and $\vect{\beta}_\mathrm{s}$, with each element being either zero or one and the boundary condition
\begin{equation}
(\vect{\beta}_\mathrm{p}+\vect{\beta}_\mathrm{s})\cdot(\vect{\beta}_\mathrm{p}+\vect{\beta}_\mathrm{s})=n_\mathrm{species}.
\end{equation}
Here, $n_\mathrm{species}$ is the number of species in the mechanism, which is 16 for the considered case. Furthermore, the loss function during the AutoML procedure is extended by another term with hyperparameter $\beta_5$ resulting in
\begin{equation}
\mathcal{L} = \beta_1 \press{L_\mathrm{adv.}} + \beta_2 L_\mathrm{pixel} + \beta_3 L_\mathrm{gradient} + \beta_4 L_\mathrm{species} + \beta_5 L_\mathrm{time}. 
\end{equation}
The new term $\beta_5 L_\mathrm{time}$ measures "a posteriori" accuracy by considering the MSE for multiple successive time steps compared to the result achieved with the full PIESRGAN. This term can become very expensive. Therefore, it was evaluated on a very small toy problem with $32^3$ cells instead of the full domain. Five time steps with fixed time step size were considered, and $\beta_5$ was chosen of the order of $\beta_4$ while reducing $\beta_2$ to remain at $\sum_i \beta_i = 1$.

Overall, the procedure to run efficient PIESRGAN-LES can be summarized by the following steps:
\begin{enumerate}
\item Run suitable DNSs to generate a database $\Phi_\mathrm{H}$.
\item Compute $\Phi_\mathrm{H}/\Phi_\mathrm{F}$ data pairs for the training of the network.
\item Train the full PIESRGAN.
\item Optimize the full PIESRGAN to a reduced PIESRGAN using AutoML.
\item Use the reduced PIESRGAN to study the target problem with PIESRGAN-LES.
\end{enumerate}

For the present case, the set of CH$_4$, CO, OH, and CO$_2$ as primary species was found to give very good results. This is interesting for multiple reasons. First,  neither $\mathrm{H}_2\mathrm{O}$ nor $\mathrm{H}^\circ/\mathrm{H}_2$ are included in the set of primary species, even though they are known to be important for methane oxidation as product of the complete oxidation and in crucial elementary reactions, such as $\mathrm{CH}_4 + \mathrm{H}^\circ \rightarrow \mathrm{CH}_3^\circ + \mathrm{H}_2$ and $\mathrm{H}^\circ +\mathrm{O}_2 \rightarrow \mathrm{OH}^\circ + \mathrm{O}^\circ$, respectively. Second, it emphasizes the importance of the CO consumption elementary reaction, $\mathrm{CO} + \mathrm{OH}^\circ \rightarrow \mathrm{CO}_2 + \mathrm{H}^\circ$, which is the key reaction affecting CO emissions in the post flame region. Note that the sets of primary and secondary species do not indicate which species are physically the most relevant. Instead, these sets quantify the level of accuracy to which a species needs to be known during a simulation.

\subsection{A priori testing}
\begin{figure*}[!htb]
    \picsize
    \centering
        \begin{subfigure}[b]{40mm}
            \centering
            \includegraphics[width=\textwidth]{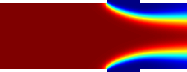}
            \vspace*{-10mm}
            \caption*{\colorbox{white}{$Z_{\mathrm{H}}$}}    
            \label{dpl:sfig:recon:HZ}
        \end{subfigure}
        \hspace{-1mm}
        \begin{subfigure}[b]{40mm}
            \centering
            \includegraphics[width=\textwidth]{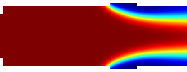}
            \vspace*{-10mm}
            \caption*{\colorbox{white}{$Z_{\mathrm{F}}$}}   
            \label{dpl:sfig:recon:FZ}
        \end{subfigure}
        \hspace{-1mm}
        \begin{subfigure}[b]{40mm}
            \centering
            \includegraphics[width=\textwidth]{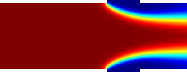}
            \vspace*{-10mm}
            \caption*{\colorbox{white}{$Z_{\mathrm{R}}$}}   
            \label{dpl:sfig:recon:RZ}
        \end{subfigure}
                \hspace{-1mm}
        \begin{subfigure}[b]{40mm}
            \centering
            \includegraphics[width=\textwidth]{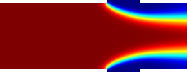}
            \vspace*{-10mm}
            \caption*{\colorbox{white}{$Z_{\mathrm{RS}}$}}   
            \label{dpl:sfig:recon:RSZ}
        \end{subfigure}
    \vskip 1mm    
    
        \begin{subfigure}[b]{40mm}
            \centering
            \includegraphics[width=\textwidth]{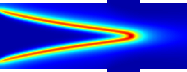}
            \vspace*{-10mm}
            \caption*{\colorbox{white}{$Y_\mathrm{CO;H}$}}    
            \label{dpl:sfig:recon:HCO}
        \end{subfigure}
        \hspace{-1mm}
        \begin{subfigure}[b]{40mm}
            \centering
            \includegraphics[width=\textwidth]{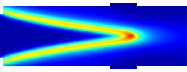}
            \vspace*{-10mm}
            \caption*{\colorbox{white}{$Y_\mathrm{CO;F}$}}    
            \label{dpl:sfig:recon:FCO}
        \end{subfigure}
        \hspace{-1mm}
        \begin{subfigure}[b]{40mm}
            \centering
            \includegraphics[width=\textwidth]{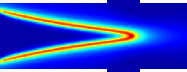}
            \vspace*{-10mm}
            \caption*{\colorbox{white}{$Y_\mathrm{CO;R}$}}    
            \label{dpl:sfig:recon:RCO}
        \end{subfigure}
                \hspace{-1mm}
        \begin{subfigure}[b]{40mm}
            \centering
            \includegraphics[width=\textwidth]{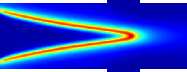}
            \vspace*{-10mm}
            \caption*{\colorbox{white}{$Y_\mathrm{CO;RS}$}}    
            \label{dpl:sfig:recon:RSCO}
        \end{subfigure}
    \vskip 1mm
   
        \begin{subfigure}[b]{40mm}
            \centering
            \includegraphics[width=\textwidth]{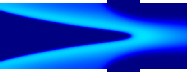}
            \vspace*{-10mm}
            \caption*{\colorbox{white}{$Y_\mathrm{OH;H}$}}    
            \label{dpl:sfig:recon:HOH}
        \end{subfigure}
        \hspace{-1mm}
        \begin{subfigure}[b]{40mm}
            \centering
            \includegraphics[width=\textwidth]{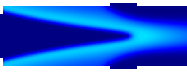}
            \vspace*{-10mm}
            \caption*{\colorbox{white}{$Y_\mathrm{OH;F}$}}    
            \label{dpl:sfig:recon:FOH}
        \end{subfigure}
        \hspace{-1mm}
        \begin{subfigure}[b]{40mm}
            \centering
            \includegraphics[width=\textwidth]{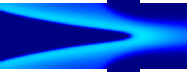}
            \vspace*{-10mm}
            \caption*{\colorbox{white}{$Y_\mathrm{OH;R}$}}    
            \label{dpl:sfig:recon:ROH}
        \end{subfigure}
                \hspace{-1mm}
        \begin{subfigure}[b]{40mm}
            \centering
            \includegraphics[width=\textwidth]{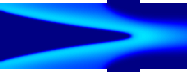}
            \vspace*{-10mm}
            \caption*{\colorbox{white}{$Y_\mathrm{OH;RS}$}}    
            \label{dpl:sfig:recon:RSOH}
        \end{subfigure}
    \vskip 1mm   
    
        \begin{subfigure}[b]{40mm}
            \centering
            \includegraphics[width=\textwidth]{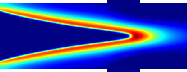}
            \vspace*{-10mm}
            \caption*{\colorbox{white}{$Y_\mathrm{H;H}$}}    
            \label{dpl:sfig:recon:HH}
        \end{subfigure}
        \hspace{-1mm}
        \begin{subfigure}[b]{40mm}
            \centering
            \includegraphics[width=\textwidth]{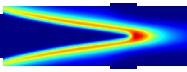}
            \vspace*{-10mm}
            \caption*{\colorbox{white}{$Y_\mathrm{H;F}$}}    
            \label{dpl:sfig:recon:FH}
        \end{subfigure}
        \hspace{-1mm}
        \begin{subfigure}[b]{40mm}
            \centering
            \includegraphics[width=\textwidth]{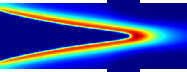}
            \vspace*{-10mm}
            \caption*{\colorbox{white}{$Y_\mathrm{H;R}$}}    
            \label{dpl:sfig:recon:RH}
        \end{subfigure}
                \hspace{-1mm}
        \begin{subfigure}[b]{40mm}
            \centering
            \includegraphics[width=\textwidth]{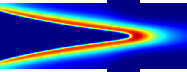}
            \vspace*{-10mm}
            \caption*{\colorbox{white}{$Y_\mathrm{H;RS}$}}    
            \label{dpl:sfig:recon:RSH}
        \end{subfigure}
    \vskip 1mm   
    
        \begin{subfigure}[b]{40mm}
            \centering
            \includegraphics[width=\textwidth]{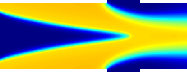}
            \vspace*{-10mm}
            \caption*{\colorbox{white}{$Y_\mathrm{H2O;H}$}}    
            \label{dpl:sfig:recon:HH2O}
        \end{subfigure}
        \hspace{-1mm}
        \begin{subfigure}[b]{40mm}
            \centering
            \includegraphics[width=\textwidth]{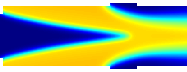}
            \vspace*{-10mm}
            \caption*{\colorbox{white}{$Y_\mathrm{H2O;F}$}}    
            \label{dpl:sfig:recon:FH2O}
        \end{subfigure}
        \hspace{-1mm}
        \begin{subfigure}[b]{40mm}
            \centering
            \includegraphics[width=\textwidth]{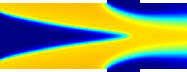}
            \vspace*{-10mm}
            \caption*{\colorbox{white}{$Y_\mathrm{H2O;R}$}}    
            \label{dpl:sfig:recon:RH2O}
        \end{subfigure}
                \hspace{-1mm}
        \begin{subfigure}[b]{40mm}
            \centering
            \includegraphics[width=\textwidth]{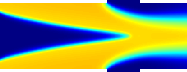}
            \vspace*{-10mm}
            \caption*{\colorbox{white}{$Y_\mathrm{H2O;RS}$}}    
            \label{dpl:sfig:recon:RSH2O}
        \end{subfigure}
    \vskip 3mm   
       \caption{Visualization of DNS (first column), filtered (second column), and reconstructed fields for the $V_\mathrm{inlet}=\SI{2.0}{\meter\per\second}$ case employing PIESRGAN (third column) and $\mathrm{PIESRGAN}_\mathrm{S}$ (fourth column). Results for mixture fraction, CO mass fraction (primary species), OH mass fraction (primary species), $\mathrm{H}^\circ$ mass fraction (secondary species), and $\mathrm{H}_2\mathrm{O}$ mass fraction (secondary species) are shown.}
        \label{dpl:fig:recon}
    \end{figure*}

The accuracy of the PIESRGAN model for lean premixed combustion super-resolution is first evaluated based on an a priori test for the full model (denoted "R") and the reduced PIESRGAN, PIESRGAN$_\mathrm{S}$ (denoted "RS"). For that, the networks were trained with the 
$V_\mathrm{inlet}=\SI{0.0}{\meter\per\second}$ and $V_\mathrm{inlet}=\SI{4.0}{\meter\per\second}$ case data and applied to reconstruct fields of the $V_\mathrm{inlet}=\SI{2.0}{\meter\per\second}$ case containing the reaction zone. The results are shown in \reff{dpl:fig:recon}. The visual agreement between fully resolved data and reconstructed data is very good. The filtered data clearly lack resolution around the reaction zone, and it cannot be expected that it is possible to accurately advance the filtered fields without an advanced closure model. The agreement between the full and reduced PIERSGAN model is also striking, suggesting that the combination of super-resolved full transport equations for some important species with a precomputed network look-up for other species is a promising approach. Note that only the CO mass fraction field was directly reconstructed by the PIESRGAN-networks. Temperature and mixture fraction are computed from all reconstructed mass fraction fields and are only shown for demonstration purposes here.

\subsection{A posteriori testing}
\begin{figure}[!htb]
\picsize
	\centering
    \picbox{\input{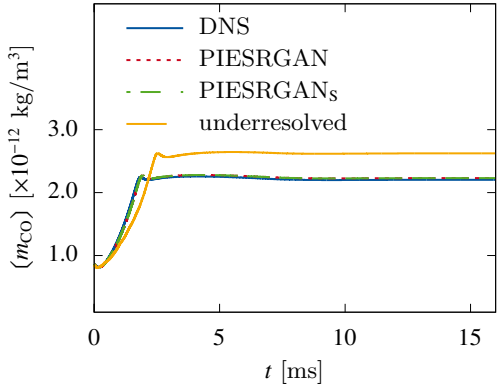}}
    \vskip 1mm
	\caption{Plot of the domain-averaged CO mass over time for a fully resolved case, a full PIESRGAN case, a reduced PIESRGAN case, and an underresolved case without closure for the $V_\mathrm{inlet}=\SI{2.0}{\meter\per\second}$ case.}
	\label{dpl:fig:conservation}
\end{figure}

For a posteriori testing, the $V_\mathrm{inlet}=\SI{2.0}{\meter\per\second}$ case was recomputed on an eight times coarser mesh in each non-periodic direction using PIESRGAN and PIESRGAN$_\mathrm{S}$ closures. All cases were initialized with a 1-D flamelet solution and zero initial velocity. In \reff{dpl:fig:conservation}, the domain-averaged CO mass $\avgst{m_\mathrm{CO}}$ over time is shown for the fully resolved DNS case, the PIESRGAN case, and an underresolved case without closure. The agreement between the fully resolved case and PIESRGAN is very good for all time steps. The underresolved case features a smaller increase of CO mass in the beginning and is not able to accurately predict CO consumption towards steady-state.  Also in this a posteriori test, the reduced PIERSGAN model delivers results extremely close to the full model.

\subsection{Generality}
\begin{figure}[!htb]
\picsize
	\centering
    \picbox{\input{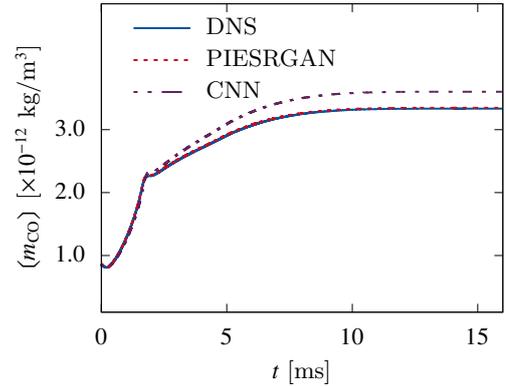}}
    \vskip 1mm
	\caption{Plot of the domain-averaged CO mass over time for a fully resolved case, a PIESRGAN case, and CNN case for the $V_\mathrm{inlet}=\SI{8.0}{\meter\per\second}$ case. Note that the reduced PIESRGAN gave very similar results to the full PIESRGAN and was therefore skipped in this plot.}
	\label{dpl:fig:extrapolation}
\end{figure}

The a posteriori test showed that it is possible to accurately predict the $V_\mathrm{inlet}=\SI{2.0}{\meter\per\second}$ case with a network trained with $V_\mathrm{inlet}=\SI{0.0}{\meter\per\second}$ and $V_\mathrm{inlet}=\SI{4.0}{\meter\per\second}$ case data. This was a classical "interpolation" case. However, an important question is how general the model is and whether the network has certain "extrapolation" abilities, i.\,e., can predict a case outside of the training range. For this, the $V_\mathrm{inlet}=\SI{8.0}{\meter\per\second}$ case is rerun on a coarser mesh. This case features higher velocities and consecutive strain rates, which - as explained earlier - implicitly effect also the chemistry fields and were not seen during training. The results are presented in \reff{dpl:fig:extrapolation} for the average CO mass in the domain. As before, the agreement between the fully resolved case and PIESRGAN is very good. 

Besides the fully resolved result and the PIESRGAN result, a result for a network trained without the discriminator part (denoted "CNN") is given in \reff{dpl:fig:extrapolation}. The CNN case overpredicts the steady-state amount of CO mass after successfully predicting the evolution in the beginning. A detailed comparison of the PIESRGAN and CNN results reveals that the CNN has problems predicting the reactions in high velocity areas, similar to an underresolved simulation, while the PIESRGAN gives much better results there. This emphasizes the better extrapolation capabilities of GANs compared to simpler networks and is an important result. Note that this advantage is paid for by higher training cost and typically more convergence issues during training.

\section{Conclusions}
This work advances the PIESRGAN-based modeling approach to reactive finite-rate-chemistry flows. It is shown how a newly devised model, combining super-resolution with direct network lookup for chemistry computation, can be used to accurately predict the evolution of a laminar lean premixed reactive flow on a coarser filtered mesh. For that, the modeling algorithm and loss function of PIESRGAN were modified. To enable practical usage of this model, a staggered splitting in primary and secondary species is introduced, which significantly reduces computational cost and simplifies the application on current GPUs, while maintaining the accuracy almost completely.

Although a speedup was measured for the 2-D laminar target case, the usage of the model for this case seems questionable, as the DNS is cheap and fast enough to run fully resolved simulations. However, for larger 3-D laminar cases and especially for turbulent setups, this is not possible and the here developed methods, such as the splitting, allow for a significant increase in simulation speed and thus enable parameter variations, such as complex geometry and green fuels variations.
Overall, the developed approach allows for a significantly faster development of the next generation of energy devices, such as turbines with hydrogen and engines with ammonia as fuel.

The full integration of chemistry and turbulence in one PIESRGAN is demonstrated in a follow-up paper focusing on PIESRGAN for turbulent premixed flame kernels~\cite{bode2022dpt}.

\section*{Acknowledgements}
The author acknowledges computing time grants for the projects JHPC55 and TurbulenceSL by the JARA-HPC Vergabegremium provided on the JARA-HPC Partition part of the supercomputer JURECA at J\"ulich Supercomputing Centre, Forschungszentrum J\"ulich,  the Gauss Centre for Supercomputing e.V. (www.gauss-centre.eu) for funding this project by providing computing time on the GCS Supercomputer JUWELS at Jülich Supercomputing Centre (JSC), and funding from the European Union's Horizon 2020 research and innovation program under the Center of Excellence in Combustion (CoEC) project, grant agreement no. 952181. 


\bibliographystyle{elsarticle-num} 
\bibliography{literature}


%


\end{document}